\documentclass{article}

\usepackage{graphicx}
\usepackage[preprint]{neurips_2026}
\usepackage{listings}
\usepackage[utf8]{inputenc} 
\usepackage[T1]{fontenc}    
\usepackage{hyperref}       
\usepackage{url}            
\usepackage{booktabs}       
\usepackage{amsfonts}       
\usepackage{nicefrac}       
\usepackage{microtype}      
\usepackage{xcolor}         

\title{Human Decision-Making with Persuasive and Narrative LLM Explanations}

\author{%
  Laura R.~Marusich
   \\
  DEVCOM Army Research Laboratory\\
  Adelphi, MD 20783 \\
  \texttt{laura.m.cooper20.civ@army.mil} \\
  \And
  Mary Grace Kozuch Dhooghe \\
  University of Texas at Dallas \\
  Richardson, TX 75080 \\
  \And
  Jonathan Z. Bakdash \\
  University of Texas at Dallas \\
  Richardson, TX 75080 \\
  \And
  Murat Kantarcioglu \\
  Virginia Polytechnic Institute and State University\\
  Blacksburg, VA 24060 \\
}

\begin{document}

\maketitle

\begin{abstract}
  Large language models (LLMs) have the potential to aid and improve human decision-making in classification tasks, not only by providing fairly accurate predictions, but also in their ability to generate cogent narrative explanations of those predictions. Prior work has demonstrated that people generally find AI narrative explanations to be understandable, trustworthy, and convincing for changing beliefs and opinions; however, less is known about the impact of narrative explanations on objective human decision-making performance. Here we conduct a large-scale human behavioral experiment to evaluate decision-making performance with LLM-generated narrative explanations of varying persuasiveness. We found the degree of persuasiveness, or lack thereof, for LLM-based explanations did not meaningfully impact decision accuracy over a simple AI prediction alone, in agreement with typical results with explainable AI based on feature importance. We found evidence that narratives increased reliance on AI, but both when the AI prediction was correct and incorrect. Exploratory analyses also indicated that the more persuasive narratives may have had a detrimental effect on decision response times and the ability to discriminate between a correct and incorrect AI prediction. Overall, this work indicates that including narrative explanations with AI predictions may involve tradeoffs for decision-making performance, and more work is needed to determine how and when narrative explanations impact human decision-making. 
\end{abstract}

\section{Introduction}
\label{introduction}
Joint human/AI decision-making can potentially combine the strengths of AI with those of humans in a way that leads to better performance than what can be attained by either alone. There is substantial work exploring the factors that increase human trust in and reliance upon AI predictions, and to a lesser extent, factors that improve the accuracy performance of joint human/AI decisions. Recent advancements in large language models (LLMs) make interactions with AI systems increasingly intuitive and understandable to human users. For example, LLMs are very well-suited for generating narrative text explanations for AI predictions (compared to lists of decision rules, model visualizations, e.g.). They can also be prompted to generate highly persuasive arguments that can shift opinions in human users \citep{salvi_conversational_2025}.  
However, there is limited prior work on the the impact of persuasive and narrative AI explanations on objective human decision-making performance. For example, prior work with narrative explanations found increased human understanding of AI model reasoning yet no change in subjective human decisions~\cite{martens2025tell}. Human decision-making using rule/feature-based explainable AI has been widely evaluated, and it typically does not improve objective decision-making performance~\cite{schemmer2022meta,marusich_explainability}; although results may be task-dependent~\cite{leichtmann2024explainable}.

Given the growing use of narrative AI explanations, it is critical to evaluate how the persuasiveness, or lack thereof, for LLM narrative explanations impacts human reliance on AI and also objective decision-making performance. This topic reflects a key gap in the literature, and in the work described here, we are among the first to evaluate objective human decision-making with persuasive and narrative explainable AI. Our results show narrative explainable AI, regardless of persuasiveness, did not meaningfully change human decision-making accuracy over a simple AI prediction alone. However, we did find evidence for increased reliance on AI with narrative explanations for both correct and incorrect AI predictions.

In this paper, we first provide background on narrative explanations for explainable AI, persuasion and the growing body of work on the persuasive capabilities of LLMs, and the more limited research assessing \textit{joint human/AI decision-making}. Section~\ref{current} describes our hypotheses and experimental design, including the datasets and model prompts we used to generate our experimental stimuli. We report the findings from the pre-registered, large sample size behavioral experiment comparing human decision making accuracy with only AI predictions vs. AI predictions with narrative explanations of varying persuasiveness. Finally, we conclude by discussing the implications of our results, with limitations and recommendations for future work.

\section{Background and related work}
\label{background}

\subsection{Narrative AI explanations}
\citet{hartmann2022xaines} introduce narrative explanations as a framework in which AI explanations are tailored to the target audience through adaptive content, presentation style, and user feedback that can be used to iteratively improve the system.
Explingo, introduced and evaluated by \citet{zytek2024explingo}, employs two LLMs: a narrator, which generates narrative explanations of SHAP outputs, and a grader, which evaluates these narratives along the dimensions of accuracy, completeness, fluency, and conciseness. Their results show that LLMs can produce high-quality narrative explanations, performing particularly well in accuracy and completeness, although sometimes with reduced conciseness.

\citet{martens2025tell} create XAIStories using LLM narratives to improve user understanding of traditional AI explanation techniques. They find that for SHAP and CF explanations, general users find the narratives more convincing than the explanations alone in 93\% and 90\% of cases, respectively. Furthermore, when shown XAIStories based on SHAP instead of raw SHAP outputs, users’ objective understanding of the AI model's reasoning tended to greatly improve, but this did not impact whether the user and AI (subjective) decisions were in agreement.

\subsection{Persuasion}
\label{background:persuasion}

Social scientists have been interested in questions of persuasion for nearly a century, and it continues to be an area of active study. There are a number of widely accepted definitions and models of persuasion (see, for example, \citealt{cialdini2019influence,  okeefe2015, Petty_Cacioppo_1986, Kruglanski_Thompson_1999,Gass_Seiter_2004, Chaiken_1987}). Persuasion is distinct from other types of influence in that it is intentional, not coercive, and involves symbolic action - typically language \citep{Gass_Seiter_2022}. 

In fields such as psychology and political science, persuasion is a form of communication that intentionally aims to change or influence people's beliefs, attitudes, and/or behaviors \citep{cialdini2019influence}. In considering specifically language-based persuasion, researchers have detailed a number of features of communication that affect the persuasiveness of the message (for an overview, see \citealt{Shen_Bigsby_2013}). These features can be categorized into aspects of the \textit{content}, the \textit{structure}, and the \textit{style} of the communication.

Applying these message features to manipulate text persuasiveness is fairly intuitive for narrative-based explanations; less so for rule/feature-based explanation types. This presents an interesting opportunity to assess how narrative explanations of varying persuasiveness can affect human decision-makers' performance. It should be noted, however, that it can be difficult to produce text samples that differ purely on the dimension of persuasiveness. For example, the use of exaggerated or flowery language is a common persuasion technique \citep{cialdini2019influence} widely used in U.S. advertisements \citep{akgul2022, Amyx03052016} and often represents a natural confound. 

In the context of LLMs, persuasive ability has recently attracted significant attention, generating both excitement and concern. LLMs can match the effectiveness of human persuasion, although it depends on the content, structure, and style of communication \cite{holbling2025meta}.
%
\citet{Rogiers_Noels_Buyl_Bie_2024} survey the literature on factors affecting the persuasive capabilities of LLMs across literature, finding that multi-shot prompting is generally more effective than single-shot \citep{costello2024durably,durmus2024persuasion}, although increased engagement with static text can be similarly effective \cite{altay2023information}. This is further supported in recent work showing that in a multi-shot scenario, ChatGPT-4 is more persuasive than a human \citep{salvi_conversational_2025}.

Personalization techniques can often increase persuasiveness \citep{meguellati2024good, matz2024potential, simchon2024persuasive, Rogiers_Noels_Buyl_Bie_2024}, although \citet{hackenburg2024evaluating} find that microtargeted messaging is not more persuasive than broadly generated messages in single-shot scenarios.
Furthermore, prompts eliciting emotionally framed messaging appear to be less persuasive than those emphasizing logical reasoning explanations and non-emotional messaging \citep{simchon2024persuasive, bai2025llm}.  
However, adversarial explanations can be more persuasive than both no explanation and accurate explanations \citep{danry2025deceptive}.
Of note, prompting for differing persuasive styles produces substantial differences in output length \citep{transformer_pair}, and attempts to reduce output length through prompt constraints can inhibit persuasiveness.

Disclosure and static reminders that text is AI-generated have been shown to reduce user trust \citep{bohm2023people, karinshak2023working, bo2025rely}. To address this, \citet{bo2025rely} benchmarked interventions aimed at guiding how users rely on LLM outputs, such as general reliance disclaimers or uncertainty highlighting. They found that persistent disclaimers tend to improve appropriate reliance, whereas alternative interventions can reduce over-reliance but also increase decision time. Similarly, strategies such as appropriate personalization \citep{matz2024potential} and the incorporation of, and citation to, expert-supplied information \citep{altay2023information} can mitigate AI mistrust.

\subsection{Joint human/AI decision-making}

Accurate human involvement is essential when deploying LLMs for high-risk decision making tasks, shown in \citet{rivera2024escalation} where LLMs escalated military scenarios at a higher rate than human actors. 
\citet{marusich_uncertainty} found that providing uncertainty quantification information improves human decision making versus machine learning (ML) classifier prediction alone. 
Conversely, \citet{marusich_explainability} found that including an explanation for ML classifier predictions did not improve human decision accuracy over and above the predictions alone.
LLMs, however, introduce a distinct explanatory paradigm by generating explanations concurrently with predictions rather than through post-hoc model analysis.
LLMs can modulate explanatory persuasiveness independently of model confidence, potentially decoupling explanation conviction from prediction accuracy \citep{kalai2024calibrated}.
\citet{steyvers2025large} find that humans tend to overestimate LLM accuracy when provided with explanation when comparing human confidence with LLM provided confidence.

\citet{bo2025rely} find while testing user reliance on LLMs for LSAT logical reasoning questions, human response accuracy improves by 10\% with the assistance of LLM responses, but with experimentally controlled equal chances of correct or incorrect LLM responses, HitL accuracy is 4\% lower than LLM alone, 
motivating further study into human decision accuracy when supplied with LLM predictions of varying persuasive language styles.

Although explainable AI, especially the potential of narrative explanations, is a recent area of exploding research interest, work evaluating these techniques with human decision-makers still tends to focus on questions of trust and reliance, with little work assessing objective decision-making performance. Furthermore, when the persuasiveness of explanations are studied, it is most often as an outcome variable.
\textit{There is a gap for research that explores the effect of varying the persuasiveness of explanations upon decision-making. In this work we attempt to address these gaps.}

\section{Current Work}
\label{current}

\subsection{Design and hypotheses}

We conducted a large-scale human behavioral experiment that provided human decision-makers with LLM-generated predictions, and in some conditions, corresponding narrative explanations. In prompting the LLM to produce explanations of varying persuasiveness for this initial study, our goal was to produce the strongest possible effects, which could then be explored on a finer level in future work. We used three types of prompt (see Section \ref{models_prompts} and \ref{sec:prompt}) designed to produce neutrally-worded, somewhat persuasive, and extremely persuasive explanations, respectively. Text analyses (see \ref{sec:expl_text}) indicate that these prompts successfully produced explanations that differed across persuasion metrics, although the presence of exaggerated and manipulative language in the Extreme condition indicates this was not an entirely "pure" manipulation of persuasion, as described in Section \ref{background:persuasion}.  

We assessed not only the human reliance rates on the AI predictions, but also their decision-making performance (measured by 1: decision accuracy compared to the ground truth data, and 2: the effective calibration of participant's confidence and accuracy). In the following sections we describe our hypotheses and the methodological details of the study.  


We made the following, \href{https://aspredicted.org/6g5p-7mys.pdf}{pre-registered} hypotheses:
\begin{enumerate}
    \item Accuracy: Decision accuracy will be lower with higher persuasion. The Prediction Alone and Neutral Explanation conditions will produce similar decision accuracy, with a reduction in accuracy for Low Persuasion and an even greater reduction for Extreme Persuasion. 
    \item AI Reliance: AI reliance (the proportion of trials on which the human decision-maker follows the AI prediction) will increase with higher persuasion.
    \item Confidence Calibration: Decision confidence will be most highly calibrated (meaning a stronger positive relationship) with non-persuasive information. As in H1,  Prediction Alone and Neutral Explanation conditions will produce similar calibration levels, with decrements for the Low and Extreme Persuasion conditions.
\end{enumerate}

\subsection{Datasets}
\label{sec:dataset}

In this study we used a subset of instances from two publicly-available datasets: the {\em Census Income}, and {\em Student Performance} datasets from the UCI Machine Learning Repository~\citep{Dua:2019}. We selected these datasets because they involve real-world contexts (e.g., will a student pass or fail a class?) that are understandable and engaging for non-expert human participants. In addition, the two datasets differ in the overall accuracy prediction models can achieve in their predictions. These differences in accuracies and contexts ensure a degree of generalizability and that our findings are not limited to one specific dataset.

The original {\em Census} dataset has 48,842 instances and 14 attributes. Features such as Age, Education, and Marital Status are used to predict whether a person earns more or less than a certain income threshold (\$50,000 in 1994). We replaced any missing values in the dataset with the mode (the most frequent value for that attribute), and we adjusted all dollar amounts in the dataset for inflation, so that the prediction task was intuitive for current participants.

The {\em Student Performance} dataset has 649 instances and 33 attributes from students enrolled in a secondary school course. We averaged three of the attributes ({\em first period grade}, {\em second period grade}, and {\em final grade}) to generate a single pass/fail attribute for each instance, to be predicted based on based on features such as the parents' education level, the student's current health status, study time, etc. We used a subset of 12 of the predicting attributes in order to avoid overwhelming the human participants with information for each instance. 

For each of the two datasets, we selected a subset of 58 instances (29 of each class) that were used previously for similar behavioral experiments \citep{marusich_uncertainty}. We provided these instances to the LLM with prompts (described in detail below) to generate predictions and corresponding explanations of each type (Neutral, Low Persuasion, and Extreme Persuasion) for each instance. In the Census subset, the predictions across the three explanation condition prompts were the same for 56 out of the 58 instances. In the Student subset, 55 out of 58 were consistent. Out of the instances with consistent predictions, we chose 52 instances for our experimental stimuli (6 practice and 46 test trials) that maintained a balance between the classes and also approximated the overall LLM accuracy for the full set of 58. See the supplementary material for the complete datasets, including the generated predictions and narrative explanations.

\subsection{Models/prompts}
\label{models_prompts}
We used the OpenAI GPT-4o (Omni) model via the OpenAI API using temperature that is set to 0.1,  accessed in March 2025, to generate responses and predictions for each sample in both datasets.
For each control condition, every data instance was mapped to an identical contextual prompt (see Appendix Section~\ref{sec:prompt} for examples). The textual representation of instance-specific features was constructed in exactly the same way across all conditions. The only element that varied for a given instance across different control settings (e.g., explanation vs. persuasion) was the final instruction in the prompt. For example, the ``Low Persuasion'' variant instructed the model to “first provide a pass/fail decision, then using a persuasive style, give a brief explanation why you have this prediction.” 

For the Student Performance dataset, across all evaluated conditions, zero-shot prompting achieved prediction accuracies of 67.24\% under extreme persuasion, 68.96\% under low persuasion, and 67.41\% under the neutral condition on randomly sampled instances mentioned in Section~\ref{sec:dataset}.

For the Census dataset, zero-shot prompting similarly demonstrated stable performance across conditions, achieving accuracies ranging from approximately 72.41\% under both high and low persuasion to 75.86\% under the neutral condition on randomly sampled instances mentioned in Section~\ref{sec:dataset}.

We compared the three sets of explanations in terms of word count and reading level. While there were small differences in these distributions, the averages were fairly comparable across the different persuasion conditions, meaning that differences in decision performance between conditions are not likely to be attributable to simple differences in the amount of text or in text difficulty of the explanations. We also used text and sentiment analyses to assess whether the model prompts were effective in generating narratives that were appreciably different in terms of the metrics of persuasive text (see ~\ref{sec:expl_text} for these comparisons).

\subsection{Experimental task}
The experimental task that the participants completed was developed and hosted using the jsPsych JavaScript library \cite{de2015jspsych} and JATOS (Just Another Tool for Online Studies; \citealp{Lange_Kühn_Filevich_2015}). In all conditions, participants viewed a tabular description of an individual, an AI prediction for the classification of that individual (e.g., Pass or Fail), and a two-alternative forced choice for the participant's response. See \ref{sec:screenshot} for a screenshot of an example trial.

In the explanation conditions, a narrative explanation also appeared after the AI prediction. This explanation text did not appear all at once, but rather word-by-word (similar to a Typewriter Effect). The interval was randomly selected between 0ms (75\% of the time) and 250ms (25\% of the time). This was done to encourage participants to read the explanation text instead of skimming or skipping it altogether, and also to mimic the the appearance of text in typical LLM chatbots. When the narrative explanation was fully ``generated'', the response options appeared, and the timer for response times began. For the prediction alone condition, the response time started when the prediction was displayed. In all conditions, the timer ended when the participant clicked one of the two response buttons. After this response, participants then entered their confidence in that choice, on a Likert scale of 1 (No Confidence) to 7 (Full Confidence). Feedback was then displayed, indicating whether or not the previous choice was correct.

\begin{table}[t]
\caption{Participant characteristics.}
\label{table:participants} 
\begin{center}
\begin{tabular}{l r r l r r} 
\toprule
 Characteristic & n & Percent \% & Characteristic & n & Percent \%\\ 
 \midrule
 Sex & & & Education\\ 
\hspace{3mm}Male & 158 & 49.38 &  \hspace{3mm}Less than HS diploma & 2 & 0.63\\
 \hspace{3mm}Female & 158 & 49.38 & \hspace{3mm}HS diploma/equivalent & 49 & 15.31\\
 \hspace{3mm}Prefer not to answer & 4 & 1.25 & \hspace{3mm}Some college but no degree & 62 & 19.38\\
  Age & & &  \hspace{3mm}Associate's degree & 43 & 13.44\\
 \hspace{3mm}18-24 & 12 & 3.75 &  \hspace{3mm}Bachelor's degree & 108 & 33.75\\
 \hspace{3mm}25-34 & 88 & 27.50 & \hspace{3mm}Graduate degree & 56 & 17.50\\
 \hspace{3mm}35-44 & 97 & 30.31\\
 \hspace{3mm}45-54 & 80 & 25.00\\
 \hspace{3mm}55-64 & 24 & 7.50\\
 \hspace{3mm}65+ & 19 & 5.94\\
 \bottomrule
\end{tabular}
\end{center}
\vskip -0.1in
\end{table}

\subsection{Procedure}
We recruited 40 participants in each of 8 experimental conditions, for a total of 320 participants (see Table \ref{table:participants} for demographic information). These participants were recruited from the online platform \href{https://www.prolific.co}{Prolific}. They received \$5.00 for their participation in the experiment, and the majority completed the task within 20 minutes (i.e., well above the U.S. federal minimum hourly wage). This research received Institutional Review Board (IRB) approval.

After providing informed consent, participants viewed a series of instructional screens with examples. They completed 6 practice trials that were the same for all participants in the same dataset condition. They then completed 20 test trials. For each participant this set of 20 instances was chosen from the larger set of 46 randomly with the constraint that the correct response values were balanced (e.g. 10 trials with "pass" and 10 with "fail" as the correct response). This allowed for substantial variations in AI accuracy for each participant. After the 20 test trials, participants provided responses to a brief series of questionnaires, including demographics, self-reported strategies, subjective usability and persuasiveness ratings, AI attitudes, and an assessment of risk literacy~\citep{cokely2012measuring}; see supplementary material.  

\subsection{Confirmatory results}

Before data analysis, we conducted outlier removal by excluding individual trials with response times that exceeded three standard deviations above the participant's mean response time. In total, 145 out of 6400 trials (2.3\%) were excluded from the dataset.

\subsubsection{Accuracy}

\begin{figure}
  \centering
 \includegraphics[width=\textwidth]{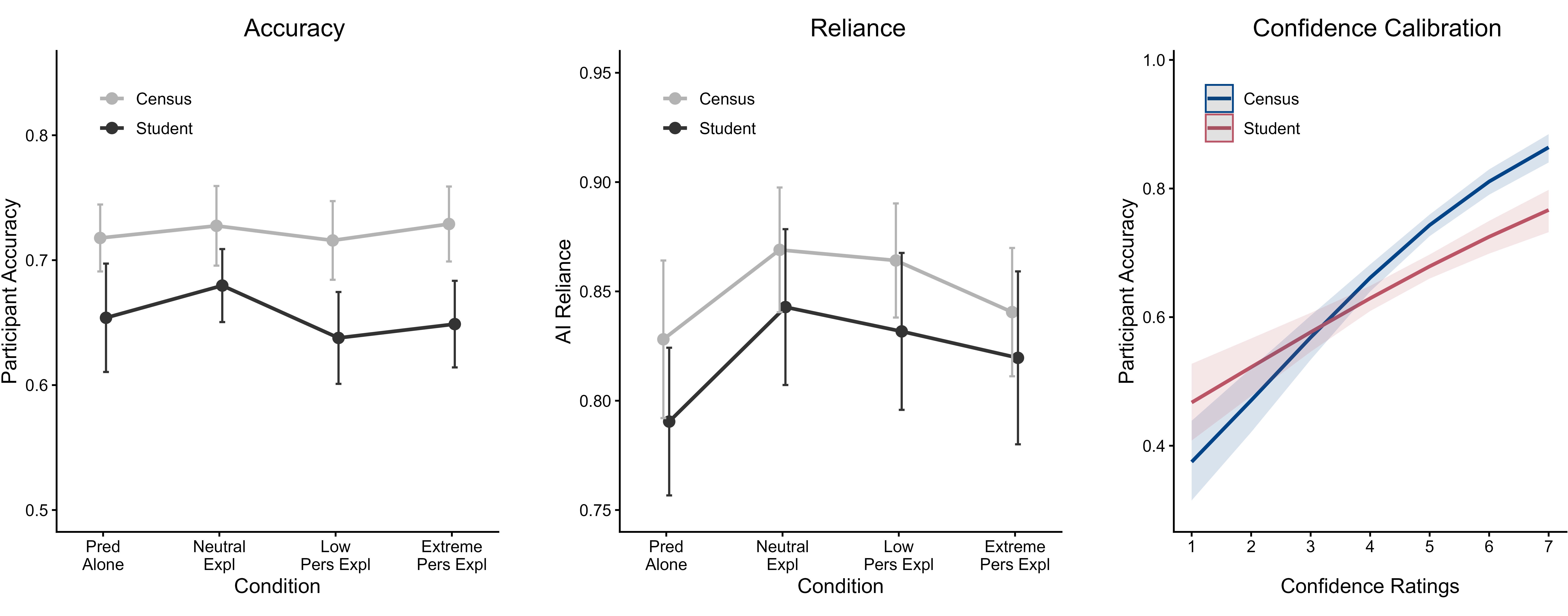}
  \caption{Left: Average participant accuracy across the two dataset conditions and four explanation conditions. Center: Average participant reliance rate across the two dataset conditions and four explanation conditions. Right: Predicted effects (level 2/overall results of multilevel model) of confidence ratings and dataset upon accuracy. Steeper, positively-sloped lines indicate better confidence calibration. Across all plots, error bars and shaded areas represent 95\% confidence intervals.}
  \label{confirmatory_figs}
\end{figure}

An omnibus 4 (Explanation Condition) x 2 (Dataset) ANOVA for
mean participant accuracy indicated that only the main effect for Dataset was significant ($F(1,312) = 33.42, {p} < 0.001, \eta_p^2 = 0.097$), with accuracy higher for the Census dataset than the Student dataset (see Figure \ref{confirmatory_figs}, left panel). This difference is expected as the AI prediction accuracy is also lower for the Student dataset. There did not appear to be a significant change in accuracy due to differences in Explanation Condition ($F(3,312) = 0.91, {p} = 0.44, \eta_p^2 = 0.009$); nor was there a significant interaction effect between Explanation Condition and Dataset ($F(3,312) = 0.41, {p} = 0.75, \eta_p^2 = 0.004$).

\subsubsection{Reliance}

A similar analysis for participant reliance rates indicated that Explanation Condition did have a significant effect on how often participants followed the AI prediction ($F(3,312) = 3.15, {p} = 0.03, \eta_p^2 = 0.03$). Tukey's honest significance test for post-hoc comparisons indicated that reliance rates were significantly higher in the Neutral Explanation condition than in the Prediction Alone condition ($t(312) = 2.83, {p} = 0.03$). In addition, the difference between Prediction Alone and Low Persuasion Explanation conditions was marginally significant ($t(312) = 2.34, {p} < 0.09$). This reflects the pattern seen in \ref{confirmatory_figs}, center panel, in which participants in both Dataset conditions rely on the AI predictions more when they receive Neutral or Low Persuasion explanations, compared to receiving a Prediction Alone. With higher persuasion however, the reliance rate appears to taper off, indicating perhaps that participants did not find the Extreme Persuasion explanations convincing. There was also a main effect of Dataset on reliance ($F(1,312) = 6.30, {p} = 0.01, \eta_p^2 = 0.02$), with participants following the AI prediction at higher rates in the Census than in the Student dataset.

\subsubsection{Confidence calibration}

We used multilevel modeling~\citep{gelman2006data} to assess the effects of Explanation Condition and Dataset upon confidence calibration. We compared models that included as fixed-effect predictors combinations of Explanation Condition, Dataset, confidence, and their interactions, as well as varying intercepts for each participant. Model comparison indicated a positive relationship between confidence and accuracy, as well as an interaction between confidence and Dataset. As depicted in Figure \ref{confirmatory_figs}, right panel, confidence calibration is better in the Census dataset condition (meaning that participants reported higher confidence as their accuracy increased) than in the Student dataset. We do not find evidence that Explanation Condition substantially impacts confidence calibration.

\subsection{Exploratory results}
\subsubsection{AI accuracy}

\begin{figure}
  \centering
 \includegraphics[width=\textwidth]{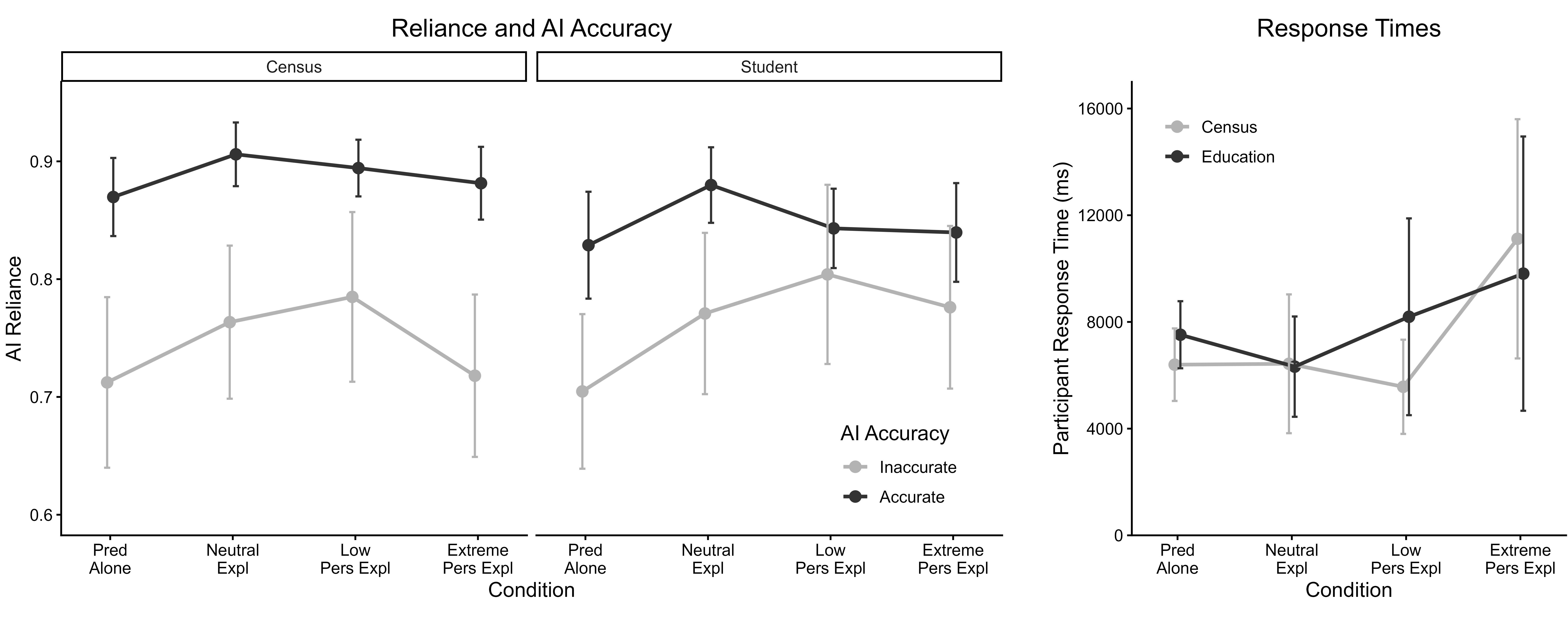}
  \caption{Left: Average participant reliance rate on trials where the AI prediction was accurate vs. inaccurate, across the two dataset conditions and four explanation conditions. Right: Average participant response times (in milliseconds) across the two dataset conditions and four explanation conditions.}
  \label{exploratory_figs}
\end{figure}

We conducted additional exploratory analyses to better understand how AI reliance rates differed by Explanation Condition without corresponding changes in overall decision accuracy or confidence calibration. We first explored participants' reliance rates on trials where the AI prediction was correct, compared to those where it was incorrect. As depicted in Figure \ref{exploratory_figs}, left panel, reliance was higher when the AI prediction was correct than when it was incorrect, indicating a level of discernment in the participants about the AI's accuracy, especially in the Prediction Alone condition. In the explanation conditions, however, the increasing reliance rates appear to be driven substantially by incorrect prediction trials, suggesting reduced discernment in explanation conditions and perhaps explaining why human accuracy does not meaningfully change across conditions. 

\subsubsection{Response times}

Another measure of decision-making performance that can be informative is the time it takes to make a response. As a reminder, response times measured the time from the appearance of the prediction (Prediction Alone) or the complete appearance of the word-by-word explanation (Explanation conditions) to the participant response. 
An omnibus ANOVA for response time performance revealed a main effect of Explanation Condition upon response time ($F(3,312) = 3.00, {p} = 0.03, \eta_p^2 = 0.03$), with pairwise tests indicating this effect is driven by the much slower times in the Extreme Persuasion condition (See Figure \ref{exploratory_figs}, right panel). As demonstrated in the very wide confidence intervals, there is also a great deal of variability in response time in the persuasion conditions. 

\subsubsection{Outperforming AI prediction accuracy}

Often the goal of augmenting human decision-making with AI predictions is to enable accuracy performance that is better than either the human or the AI accuracy alone. Previous work with similar decision-making tasks (e.g., \citealt{marusich_explainability, marusich_uncertainty})
demonstrates that access to AI prediction information does tend to substantially increase accuracy over human-alone baselines. Outperforming the baseline AI accuracy is not as common, but it is another metric that can be used to assess the impact of different AI explanation conditions. Exploratory analyses indicate that a nontrivial number of participants outperformed the AI prediction accuracies (see Table \ref{table:outperformAI}). Interestingly, the rates of outperforming the AI appear to be similar in the Prediction Alone and Neutral Explanation conditions, and similarly lower in the Extreme and Low Persuasion Explanation conditions. The difference in proportions is not statistically significant. However, in combination with the pattern of results above, this adds to the evidence that explanations engineered to be persuasive do not seem to be beneficial for improving human decision-making performance. 

\begin{table}[h]
\caption{Number of participants who outperformed the AI}
\label{table:outperformAI} 
\begin{center}
\begin{small}
\begin{tabular}{r r c r} 
\toprule
&  \multicolumn{3}{c}{Human Accuracy}\\
Explanation Condition & \textgreater AI (\textit{n}) & $\le$AI (\textit{n}) &\textgreater AI (\%)\\  
 \midrule
Prediction Alone & 24 & 56 & 30.0\%\\
 Neutral & 25 & 55 & 31.2\%\\
 Low Persuasion & 18 & 62 & 22.5\%\\
 Extreme Persuasion & 19 & 61 & 23.8\%\\
\bottomrule
\end{tabular}
\end{small}
\end{center}
\vskip -0.1in
\end{table}

\section{General discussion}
\label{discussion}
Our results indicate that providing LLM prediction explanations of varying persuasiveness may not provide a meaningful benefit to average human decision-making accuracy or confidence calibration. This is consistent with a meta-analysis showing that feature/rule-based AI explanations often do not improve decision-making performance over and above AI predictions alone~\cite{schemmer2022meta}. The LLM explanations did, however, appear to increase average reliance on the AI predictions. 


Our exploratory analyses focused on understanding why the higher AI reliance rates in the explanation conditions did not translate to higher accuracy rates, as might be expected with fairly high AI prediction accuracy. We found that participants in the Prediction Alone condition demonstrated a fair degree of discernment about the AI prediction accuracy; their reliance rates were higher for trials where the AI prediction was accurate than for trials where the prediction was inaccurate. In the explanation conditions, while overall reliance rates increased, this discernment was not maintained to the same degree. Reliance rates for inaccurate AI predictions increased as much as, and in some cases more than, the reliance rates for accurate AI predictions. This is somewhat counterintuitive, but a human simply following a system without or with only minimal discernment is not a new challenge~\cite{bainbridge1983ironies}.  

In addition, we found that response times (after information was presented to the decision) increased in length and variability with persuasive explanations. While average decision accuracy was not affected by persuasive explanations, it appears that they did cause large fluctuations in the time to make each decision. In time-sensitive contexts/applications, response time can be as critical a performance metric as decision accuracy. 

Finally, our findings suggested that fewer participants in the persuasive explanation conditions were able to outperform the AI prediction accuracy than those in the Prediction Alone and Neutral Explanation conditions. Taken together, these results indicate that providing persuasive arguments for an AI prediction aid is not likely to be helpful for a decision-makers' performance, and in some respects it can be potentially detrimental. 

\section{Limitations and future work} 
\label{limitations}

Here we describe a number of limitations in the current work, as well as opportunities for further study. For example, the combination of the rapid development of LLMs with the time-consuming nature of conducting human behavioral experiments meant that newer models had already appeared by the time our data collection was complete. Although our findings are based on stimuli generated by a less current model, we have also generated narrative explanations using the same prompts with more recent models. Based on comparisons of prediction accuracy and other text analyses between our stimuli and newer outputs (see \ref{sec:recent_models}), we do not expect that conducting human experiments with these newer models would lead to substantially different patterns of results, but it is certainly a question worth exploring, both with more recent and with different families of LLM. 

In addition, although the task here involved repeated judgments for each participant, the persuasion itself was one-shot, as opposed to interactive or involving multiple attempts to persuade. The one-shot approach has benefits for experimental control and rigor (the experimental manipulation is consistent across participants with repeated 20 decisions per participant). Nevertheless, there may be additional insights gained from exploring more interactive and personalized methodologies as well as further investigation of human understanding of AI model reasoning~\cite{martens2025tell}.

Another useful avenue of future work would involve a more systematic manipulation of the various features of persuasive communication (e.g., content, structure, and style). It is worth noting that the overall effectiveness of persuasion in general, as well as specific persuasion techniques, is not a settled question in the social sciences (see ~\citealt{dellavigna2010persuasion} for a representative review). The effects of specific style elements, for example, are not uniform across the literature, and the interactive effects among and between style features and, say, content features are also not well-understood. 

We also note that the work described here is focused on good-faith persuasive narrative explanations, meaning that the explanations were aligned with the actual AI predictions. Unexplored here is the world of adversarial persuasive explanations, where the aim of the explanation is to deceptively or manipulatively influence the decision-maker's choice. This line of research can potentially deepen the understanding of persuasive LLM explanations for joint human/AI decision-making, but it involves a host of associated ethical issues that should not be neglected. 

\section{Conclusion}
\label{conclusion}

This work represents one of the first attempts to assess the effects of LLM-generated narrative explanations of varying persuasiveness upon objective human decision-making. This type of large-scale behavioral experiment is challenging but critical to understand the practical implications of joint human/AI decision-making in real world tasks. For example, interventions that influence the degree of human trust, or other subjective measures such as user experience, in AI systems may be of limited practical relevance if there is no associated impact to the ultimate decisions being made or the accuracy of those decisions. Our work indicates that providing narrative LLM-based explanations, while somewhat increasing the reliance rate upon AI predictions, did not substantially affect overall decision-making accuracy in comparison to AI predictions alone, no matter the degree of explanation persuasiveness. We found some initial evidence that more persuasive narratives have diminishing returns or even detrimental effects upon decision-making ability, but we emphasize that there is ample opportunity for further research into this area to fully understand the tradeoffs associated with narrative explanations and persuasive text for joint human/AI decision-making. Furthermore, the results indicate narrative AI explanations do not necessarily improve actual human decision-making.

\begin{ack}
The views and conclusions contained in this document are those of the authors and should not
be interpreted as representing the official policies, either expressed or implied, of the U.S. DEVCOM Army Research Laboratory or the U.S. Government. The U.S. Government is authorized to reproduce and distribute reprints for Government purposes notwithstanding any copyright notation

\end{ack}

\bibliographystyle{apalike}
\bibliography{main}


\appendix

\section{Technical appendices and supplementary material}

\lstset{
  basicstyle=\ttfamily\small,
  frame=single,
  breaklines=true,
  columns=fullflexible
}

\subsection{LLM prompt examples}
\label{sec:prompt}

In Sections \ref{sec:prompt:edu} and \ref{sec:prompt:census} we provide examples of the prompts used to create narrative explanations for each of the three explanation conditions using each of the two datasets. In Sections \ref{sec:response:edu} and \ref{sec:response:census} we provide examples of the prompts used to create narrative explanations for each of the three explanation conditions using each of the two datasets. 

\subsubsection{Prompt examples for Student Performance dataset}\label{sec:prompt:edu}

\begin{lstlisting}[caption={ Neutral Explanation Prompt for Student Performance Dataset}, label={lst:edu:exp}]
 Given a student who is in high school, whose Age is 17, mother's education level is High school (completed),  father's education level is High school (completed), Father's job is Other,  reason for selecting the school is Close to home, time spent studying each week  is 5-10 hours,  number classes the student has failed previously is 0,  question to: does the student want to go to higher education (university)? is yes,  quality of family relationships is Moderate,  weekend alcohol Consumption is Very low,  current health status is Good,  number of school absences is 2,  predict whether this student will pass or fail the class. First, please just say pass or fail, then give an explanation why you have this prediction. 
 \end{lstlisting}

\begin{lstlisting}[caption={Low Persuasive Prompt for Student Performance Dataset}, label={lst:edu:pers}]
Given a student who is in high school, whose Age is 17, mother's education level is High school (completed),  father's education level is High school (completed), Father's job is Other,  reason for selecting the school is Close to home, time spent studying each week  is 5-10 hours,  number classes the student has failed previously is 0,  question to: does the student want to go to higher education (university)? is yes,  quality of family relationships is Moderate,  weekend alcohol Consumption is Very low,  current health status is Good,  number of school absences is 2,  predict whether this student will pass or fail the class. First, please just say pass or fail, then using a persuasive style, give a brief explanation why you have this prediction. 
\end{lstlisting}
\begin{lstlisting}[caption={Extreme Persuasive Prompt for Student Performance Dataset}, label={lst:edu:propoganda}]
 Given a student who is in high school, whose Age is 17, mother's education level is High school (completed),  father's education level is High school (completed), Father's job is Other,  reason for selecting the school is Close to home, time spent studying each week  is 5-10 hours,  number classes the student has failed previously is 0,  question to: does the student want to go to higher education (university)? is yes,  quality of family relationships is Moderate,  weekend alcohol Consumption is Very low,  current health status is Good,  number of school absences is 2,  predict whether this student will pass or fail the class. First, please just say pass or fail, then using a propaganda and logical fallacies  style, give a brief explanation why you have this prediction.  
\end{lstlisting}

\subsubsection{Prompt examples for Census Income dataset}\label{sec:prompt:census}

\begin{lstlisting}[caption={Neutral Explanation Prompt for Census Income Dataset}, label={lst:census:exp}]
 Given an individual who is living and working in the USA: whose Age is 28, Work class (i.e., Type of employer) is Private industry (company) employee , Highest education completed is Four year college degree , Marital Status is Currently married ,  Occupation is Executive, administrative, and managerial , Relationship to the person that owns the housing unit is Wife, Race is Caucasian ,  Sex is Female, Capital Gains for the last year is 32436.816, Capital Losses for the last year is 0.0,  Hours worked per week is 28, Country of Birth is United-States.  Please predict whether this person earns more than $108,000 per year. First, please just say Yes or No, then give an explanation why you have this prediction.
\end{lstlisting}

\begin{lstlisting}[caption={Low Persuasive Prompt for Census Income Dataset}, label={lst:census:pers}]
 Given an individual who is living and working in the USA: whose Age is 28, Work class (i.e., Type of employer) is Private industry (company) employee , Highest education completed is Four year college degree , Marital Status is Currently married ,  Occupation is Executive, administrative, and managerial , Relationship to the person that owns the housing unit is Wife, Race is Caucasian ,  Sex is Female, Capital Gains for the last year is 32436.816, Capital Losses for the last year is 0.0,  Hours worked per week is 28, Country of Birth is United-States.  Please predict whether this person earns more than $108,000 per year. First, please just say Yes or No, then, using a persuasive style,  give a brief explanation why you have this prediction.
\end{lstlisting}

\begin{lstlisting}[caption={Extreme Persuasive Prompt for Census Dataset}, label={lst:census:propoganda}]
 Given an individual who is living and working in the USA: whose Age is 28, Work class (i.e., Type of employer) is Private industry (company) employee , Highest education completed is Four year college degree , Marital Status is Currently married ,  Occupation is Executive, administrative, and managerial , Relationship to the person that owns the housing unit is Wife, Race is Caucasian ,  Sex is Female, Capital Gains for the last year is 32436.816, Capital Losses for the last year is 0.0,  Hours worked per week is 28, Country of Birth is United-States.  Please predict whether this person earns more than $108,000 per year. First, please just say Yes or No, then, using a propaganda and logical fallacies  style, give a brief explanation why you have this prediction.   
\end{lstlisting}

\subsubsection{Response examples for Student Performance dataset}
\label{sec:response:edu}

\begin{lstlisting}[caption={ Neutral Explanation Response for Student Performance Dataset}, label={lst:edu:exp:resp}]
Pass.

The prediction that the student will pass the class is based on several positive indicators from the provided data. Firstly, the student has a strong academic foundation, as evidenced by having no previously failed classes. This suggests a consistent level of academic performance and understanding of the material. Additionally, the student's commitment to higher education indicates a motivation to succeed academically, which is often correlated with better performance in school.

The student's study habits, with 5-10 hours spent studying each week, are within a reasonable range for high school students and suggest a dedication to maintaining their academic responsibilities. Furthermore, the student's very low weekend alcohol consumption and good health status are positive factors that contribute to their ability to focus and perform well in school.

The moderate quality of family relationships, while not exceptional, does not appear to be a hindrance to the student's academic success. The low number of school absences (2) also indicates that the student is regularly attending classes, which is crucial for keeping up with coursework and understanding the material.

Overall, these factors collectively suggest that the student is well-positioned to pass the class.
 \end{lstlisting}

\begin{lstlisting}[caption={Low Persuasive Response for Student Performance Dataset}, label={lst:edu:pers:resp}]
Pass.

This student is poised for success in their current class, and here's why. Despite having parents with only high school education, the student is demonstrating a strong desire to pursue higher education, which is a significant motivational factor. Their commitment is further evidenced by their low number of school absences and very low weekend alcohol consumption, indicating a focus on their studies and a healthy lifestyle. The student spends a reasonable amount of time studying each week, which, coupled with their good health and moderate family relationships, creates a supportive environment for academic success. Additionally, having no history of failed classes suggests a consistent academic performance. All these factors combined point towards a high likelihood of passing the class.

\end{lstlisting}

\begin{lstlisting}[caption={ Extreme Persuasive Response for Student Performance Dataset}, label={lst:edu:propoganda:resp}]
Pass.

This student is destined to pass the class, and here's why: With a family background that values education, as both parents have completed high school, this student is already set on a path of academic success. The decision to attend a school close to home ensures that the student is not burdened by long commutes, allowing more time for study and rest. Despite the modest weekly study hours, the student's perfect record of zero failed classes speaks volumes about their dedication and capability.

Moreover, the student's aspiration to pursue higher education is a clear indicator of their commitment to academic excellence. The moderate quality of family relationships provides a stable support system, essential for maintaining focus and motivation. The very low weekend alcohol consumption is a testament to the student's discipline and prioritization of health and academics.

With only two school absences, this student demonstrates reliability and a strong attendance record, further solidifying their likelihood of passing. In conclusion, all signs point to success, and any suggestion otherwise would be a baseless attack on this student's evident potential and determination.
\end{lstlisting}

\subsubsection{Response examples for Census Income  dataset}
\label{sec:response:census}
\begin{lstlisting}[caption={ Neutral Explanation Response for Census Income Dataset}, label={lst:census:exp:resp}]
Yes.

The prediction that this individual earns more than $108,000 per year is based on several key factors:

1. **Occupation**: The individual works in an executive, administrative, and managerial role, which typically commands a higher salary range, especially in the private sector.

2. **Education**: Having completed a four-year college degree provides a strong foundation for higher earning potential, particularly in managerial positions.

3. **Capital Gains**: The individual reported significant capital gains of $32,436.816 in the last year, indicating potential investments or stock options that contribute to overall income.

4. **Work Class**: Being employed in the private industry often offers higher salaries compared to public sector roles, especially in executive positions.

5. **Marital Status**: While not directly influencing income, being married can sometimes correlate with financial stability and shared economic resources, potentially allowing for more investment opportunities.

6. **Hours Worked**: Although the individual works 28 hours per week, which is less than full-time, the nature of executive roles often includes performance-based compensation, bonuses, and stock options that can significantly boost income.

Considering these factors, it is reasonable to predict that this individual's total income, including salary and capital gains, exceeds $108,000 per year.
\end{lstlisting}

\begin{lstlisting}[caption={Low Persuasive Response for Census Income Dataset}, label={lst:census:pers:resp}]
Yes.

This individual is likely to earn more than $108,000 per year. Here's why: 

Firstly, her occupation as an executive, administrative, and managerial professional suggests she holds a high-ranking position within her company. Such roles typically command substantial salaries, especially in the private sector. Furthermore, her educational background, with a four-year college degree, aligns with the qualifications often required for such positions, enhancing her earning potential.

Additionally, her capital gains of $32,436.816 indicate a savvy approach to investments or other income-generating activities, which often correlate with higher overall income levels. Although she works 28 hours per week, which is below the traditional full-time threshold, this could imply a high hourly wage or a flexible, high-paying role that allows for reduced hours without sacrificing income.

Moreover, being married could suggest a dual-income household, potentially providing financial stability and allowing for strategic career choices that maximize income. Her demographic profile, including being Caucasian and born in the United States, may also afford her certain socio-economic advantages that can contribute to higher earnings.

In summary, her professional role, educational background, capital gains, and demographic factors collectively suggest that she is well-positioned to earn more than $108,000 annually.
\end{lstlisting}

\begin{lstlisting}[caption={ Extreme Persuasive Response for Census Income Dataset}, label={lst:census:propoganda:resp}]

Yes.

Ladies and gentlemen, let us delve into the life of this remarkable individual, a beacon of success and prosperity in the land of opportunity. She is a 28-year-old executive, a leader in the private industry, wielding her four-year college degree like a sword of triumph. Her marital status, currently married, suggests a stable and supportive personal life, which is often a cornerstone of professional success.

Now, consider her occupation: Executive, administrative, and managerial. These roles are synonymous with high earnings and significant responsibilities. She is not just a cog in the machine; she is the one steering the ship. Her race, Caucasian, and her gender, female, place her in a unique position to break barriers and shatter glass ceilings, further propelling her income potential.

Her capital gains of $32,436.816 in the last year alone are a testament to her financial acumen and investment prowess. This is not a woman who lets opportunities slip by; she seizes them with both hands. And while she works 28 hours per week, a seemingly modest number, it is a strategic choice, allowing her to balance work and life while maximizing productivity and efficiency.

Born in the United States, she is a product of the American dream, a dream that promises prosperity to those who dare to dream big and work hard. With all these factors combined, it is clear that she is not just earning more than $108,000 per year; she is thriving, a shining example of success in modern America.
\end{lstlisting}

\subsection{LLM output analysis}
\label{sec:expl_text}

\begin{figure}[h]
\centerline{\includegraphics[height=0.5\textheight]{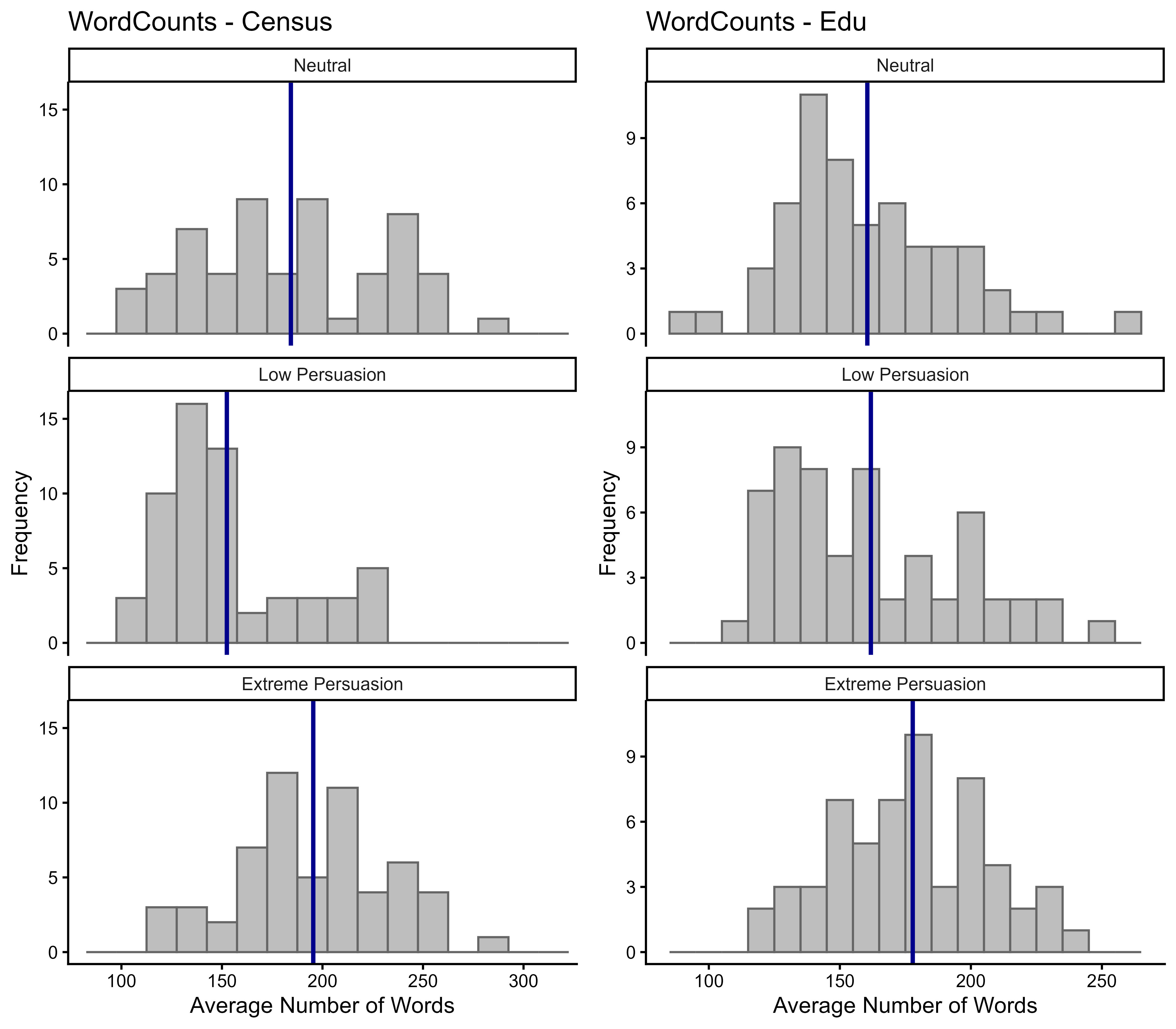}}
\caption{Distribution and mean (dark blue line) word counts for each of the three explanation types in the two dataset conditions.}
\label{fig:wordcounts}
\end{figure}

\begin{figure}[h]
\centerline{\includegraphics[height=0.4\textheight]{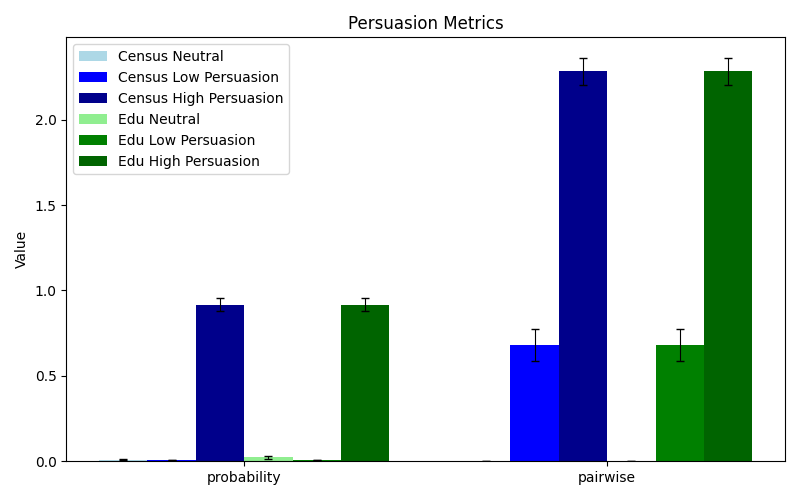}}
\caption{We measure persuasive speech using two pre-trained transformers. \textit{Probability} is measured using \protect\cite{transformer_prob}. \textit{Pairwise} is measured using \protect\cite{transformer_pair} with the \textit{Neutral} case for each record as the comparative baseline. Error bars are 95\% confidence intervals.}
\label{fig:transformers}
\end{figure}

\begin{figure}[h]
\centerline{\includegraphics[height=0.4\textheight]{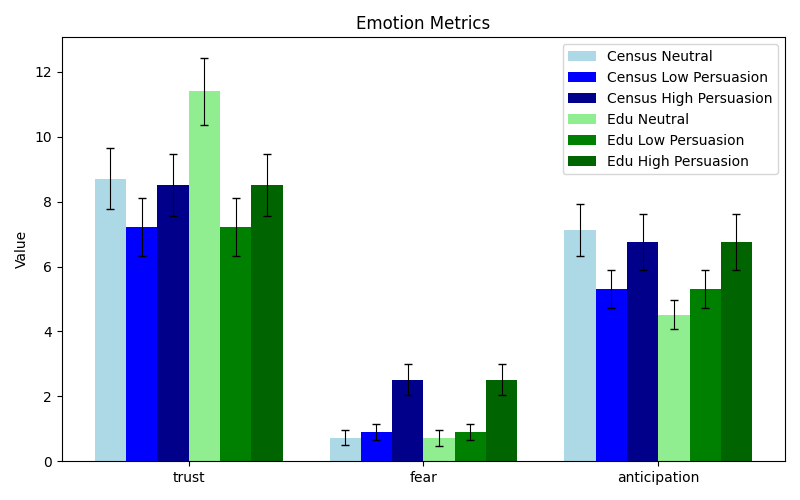}}
\caption{Emotions measured using NRCLex Raw Emotion Scores \protect\cite{emotionScores}. Error bars are 95\% confidence intervals.}
\label{fig:emotions}
\end{figure}

\begin{figure}[h]
\centerline{\includegraphics[height=0.4\textheight]{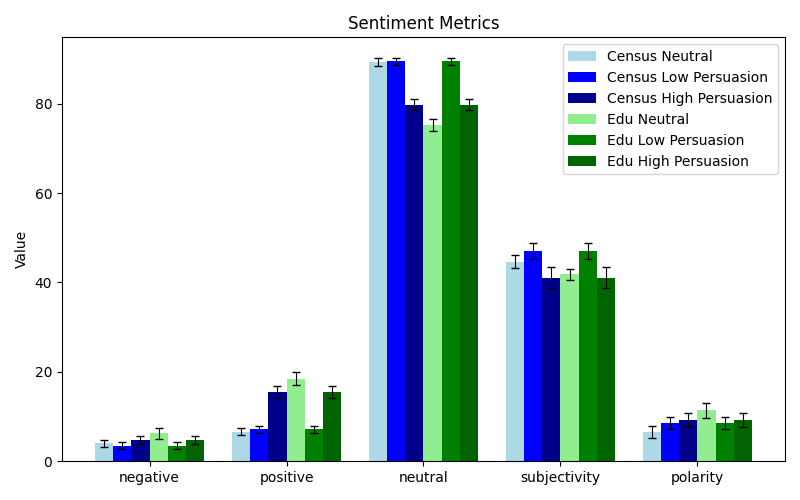}}
\caption{Sentiment is measured as percent \textit{negative, positive}, or \textit{neutral} using NLTK Sentiment Intensity Analyzer \protect\cite{NLTK, Vader}. \textit{Subjectivity} and \textit{Polarity} are measured using TextBlob Sentiment \protect\cite{TextBlobDocs}. Error bars are 95\% confidence intervals.}
\label{fig:sentiment}
\end{figure}

\begin{figure}[h]
\centerline{\includegraphics[height=0.4\textheight]{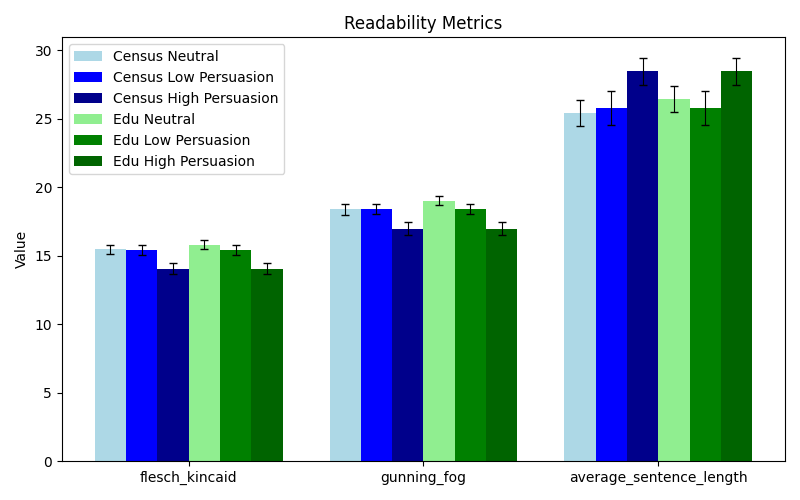}}
\caption{Readability measured using the TextStat Python library \protect\cite{TextStatDocs}. Flesch–Kincaid and Gunning Fog report readability as U.S. grade levels, where values above 12 correspond to years of college. Both metrics depend on average sentence length and average syllables per word; thus, higher scores may reflect the frequent use of multi-syllabic variable names in the datasets. Since these variable names also appear in the prompts, the prompts exhibit similarly elevated readability scores. Error bars are 95\% confidence intervals.}
\label{fig:readability}
\end{figure}

\begin{figure}[h]
\centerline{\includegraphics[height=0.4\textheight]{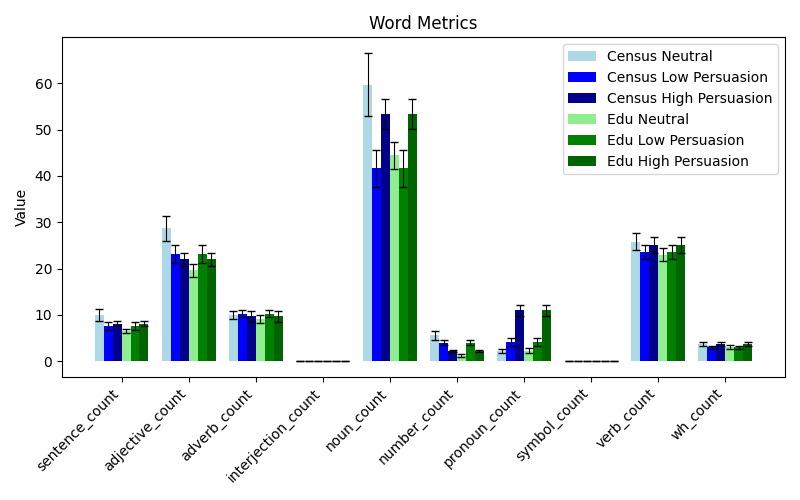}}
\caption{Average word type count collected using TextBlob Tags \protect\cite{TextBlobDocs}. \textit{wh\_count} reports average `who',`what',`where',`why', and `when' occurrences. Error bars are 95\% confidence intervals.}
\label{fig:words}
\end{figure}

\begin{figure}[h]
\centerline{\includegraphics[height=0.4\textheight]{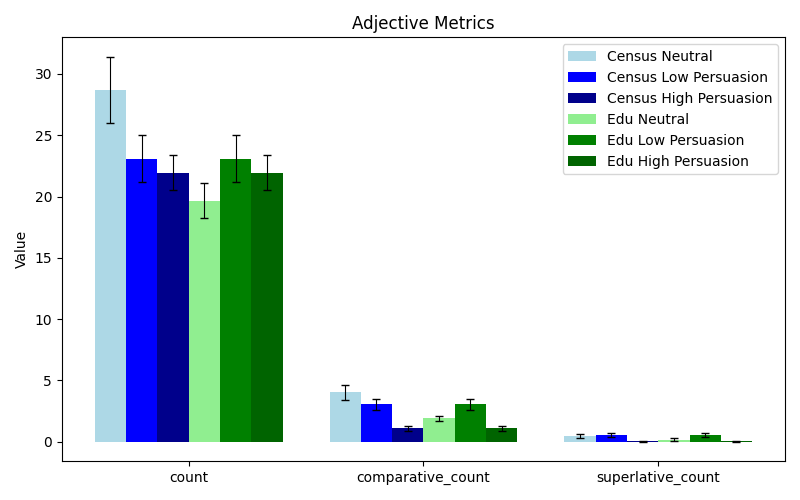}}
\caption{Average adjective type count collected using TextBlob Tags \protect\cite{TextBlobDocs}. 
\textit{count} is all adjectives while \textit{comparative\_count} is comparative adjectives such as bigger or better. Similarly, \textit{superlative\_count} is superlative adjectives such as biggest or best. Error bars are 95\% confidence intervals.}
\label{fig:adj}
\end{figure}

\begin{figure}[h]
\centerline{\includegraphics[height=0.4\textheight]{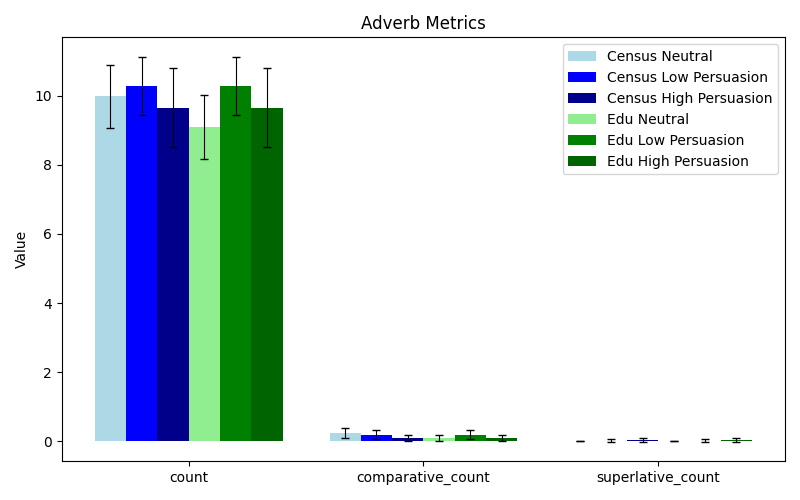}}
\caption{Average adverb type count collected using TextBlob Tags \cite{TextBlobDocs}. \textit{count} is all adverbs while \textit{comparative\_count} is comparative adverbs such as faster or more. Similarly, \textit{superlative\_count} is superlative adverbs such as fastest or most. Error bars are 95\% confidence intervals.}
\label{fig:adv}
\end{figure}

In the following analyses, including \autoref{fig:wordcounts}, \autoref{fig:transformers}, \autoref{fig:emotions}, \autoref{fig:sentiment}, \autoref{fig:readability}, \autoref{fig:words}, \autoref{fig:adj}, and \autoref{fig:adv}, we compare the text of the narrative explanations generated for each of the three persuasion conditions. We first compare the stimuli for potential confounding variables including word count or readability of the text. We then compare them on dimensions of sentiment and persuasion as a manipulation check, to ensure that our prompts successfully generated explanations of varying persuasiveness.

\clearpage
\subsection{Experimental task}
\label{sec:screenshot}
Here we provide a screenshot of an example trial, showing what the participant sees before making a response.
\begin{figure*}[h]
\begin{center}
\centerline{\includegraphics[width=\textwidth]{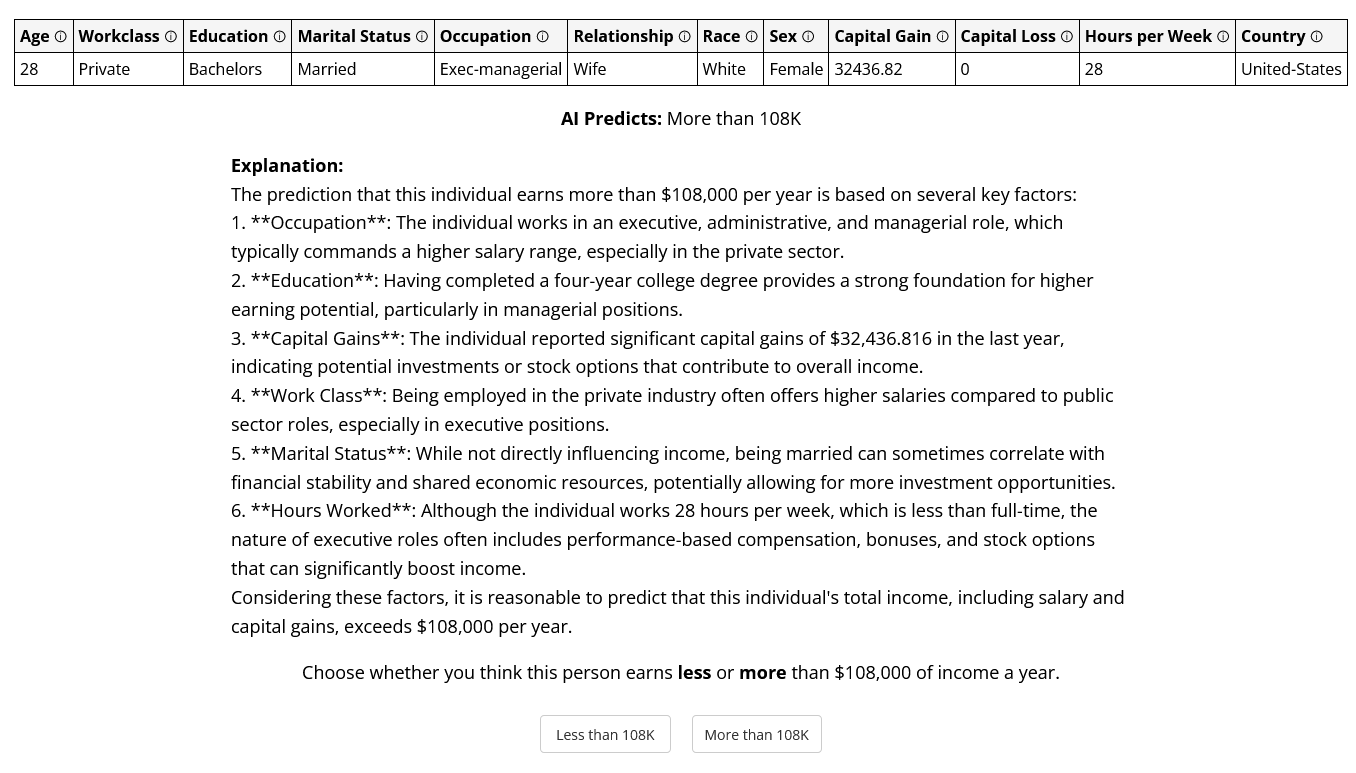}}
\caption{Example trial in the Neutral Explanation condition using the Census dataset.}
\label{screenshot}
\end{center}
\end{figure*}

\subsection{Comparison with recent models}
\label{sec:recent_models}
We compare the responses from GPT-4o, used in this work, with alternative model responses when provided with the set of prompts used in this work. For this analysis we collect responses from GPT-5.2 and GPT-5.4.

\begin{figure}
    \centering
    \includegraphics[width=\linewidth]{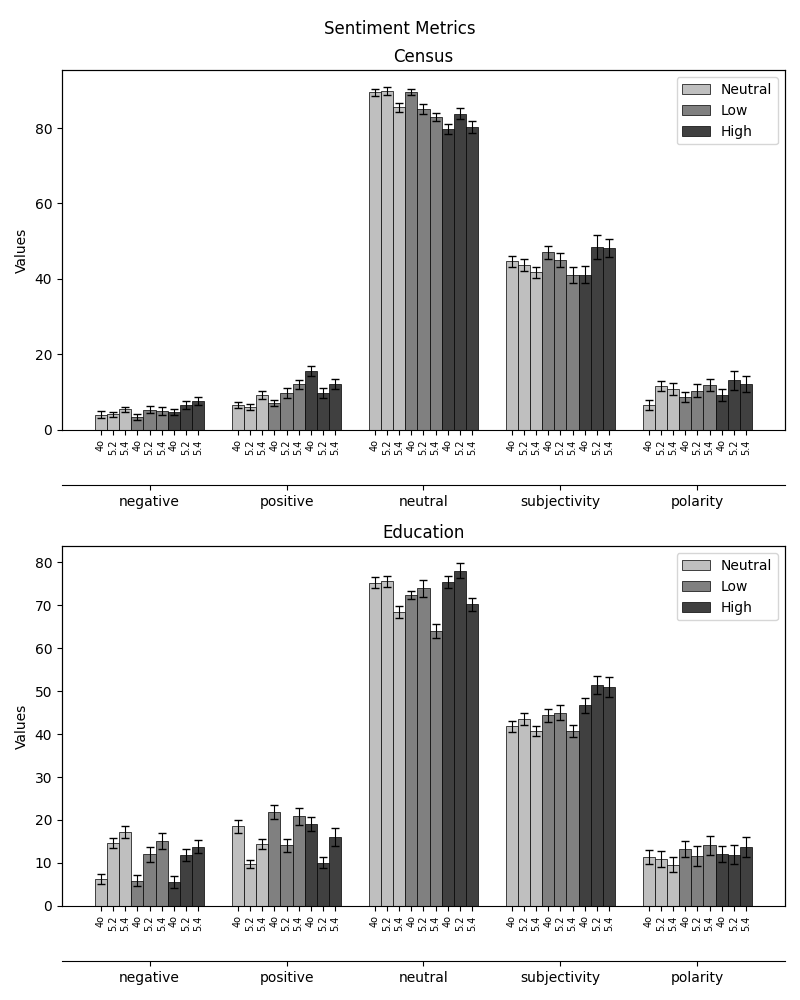}
    \caption{Sentiment analysis scores obtained from NLTK~\protect\cite{NLTK, Vader} and TextBlob~\protect\cite{TextBlobDocs} sentiment analysis tools. Error bars are 95\% confidence intervals. We report metrics for GPT-4o, GPT-5.2, and GPT-5.4 for comparison.}
    \label{fig:sentiment_metric_comp}
\end{figure}

To evaluate the linguistic and persuasive properties of the LLM outputs, we utilize NLTK’s sentiment analysis suite, which reports `Positive', `Negative', and `Neutral' textual sentiment. The choice of this metric is informed by the benchmarking framework used by \cite{potter2024hidden} in their study on the political influence of Large Language Models. 
We additionally include the sentiment scores from \cite{TextBlobDocs} including `Subjectivity' and  `Polarity' for all three models, all three persuasion manipulations, and both datasets. Results are shown in \autoref{fig:sentiment_metric_comp}. 

\begin{figure}
    \centering
    \includegraphics[width=\linewidth]{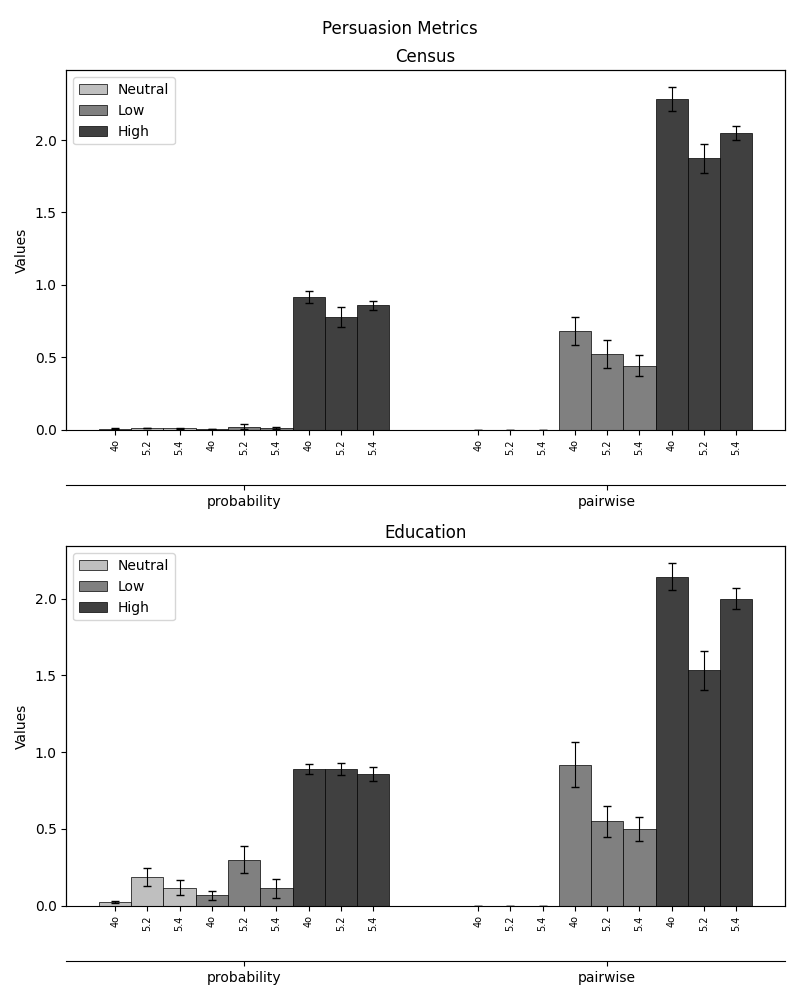}
    \caption{We measure persuasive speech using two pre-trained transformers. \textit{Probability} is measured using \protect\cite{transformer_prob} and \textit{Pairwise} measured using \protect\cite{transformer_pair} with the \textit{Neutral} case for each record as the comparative baseline. Error bars are 95\% confidence intervals. We report metrics for GPT-4o, GPT-5.2, and GPT-5.4 for comparison.}
    \label{fig:transformer_comp}
\end{figure}

We present in \autoref{fig:transformer_comp} persuasive speech levels for GPT-4o and the alternative models according to two pre-trained transformers. 
\textit{Probability} is measured using \cite{transformer_prob} which reports the probability of a text sample containing persuasive language. 
\textit{Pairwise} is measured using \cite{transformer_pair} which was trained on pairs of human annotated text pairs on a scale from [-3,3] where negative or positive values indicate that the given text sample is less or more persuasive, respectively, than the provided baseline. Annotators labeled text samples as 'marginally','moderately', or 'heavily' more or less persuasive corresponding to scores of $\pm 1$, $\pm 2$, or $\pm 3$, respectively. We utilize the \textit{Neutral} case for each record and model as the comparative baseline, all \textit{Neutral} results are therefore 0 for this metric. 

While there are differences in the magnitude of individual metrics across the different models, generally we find that the overall relationship between Persuasion Condition and model follows a similar shape.

\clearpage
\subsection{Stimuli, data, and code}

The full stimulus set, the participant data, and the analysis code are all available at \url{https://osf.io/nfht2/overview?view_only=b7b7b3d6f84f471e89fabaa7bed4a94d}.


\end{document}